\documentclass[aps,prd,twocolumn,a4paper]{revtex4-1}
\usepackage{ulem,amssymb}
\usepackage{cancel}
\usepackage{color}
\usepackage{graphicx}
\usepackage{epsfig}
\usepackage{mathrsfs}
\usepackage{amsmath,amsthm}
\usepackage[dvipsnames]{xcolor}
\usepackage{float}
\definecolor{darkblue}{RGB}{0,0,196}
\definecolor{darkgreen}{RGB}{0,120,0}
\usepackage[colorlinks=true,linktocpage=true,linkcolor=darkblue,citecolor=red,urlcolor=darkblue]{hyperref}

\newcommand{\be}{\begin{equation}}
\newcommand{\ee}{\end{equation}}
\newcommand{\ba}{\begin{eqnarray}}
\newcommand{\ea}{\end{eqnarray}}

\begin{document}

\title{Thermal dilepton production in collisional hot QCD medium in the presence of chromo-turbulent fields}

\author{Lakshmi J. Naik}
\email{jn$_$lakshmi@cb.students.amrita.edu}
\affiliation{Department of Sciences, Amrita School of Engineering, Coimbatore, Amrita Vishwa Vidyapeetham, India}

\author{V. Sreekanth}
\email{v$_$sreekanth@cb.amrita.edu}
\affiliation{Department of Sciences, Amrita School of Engineering, Coimbatore, Amrita Vishwa Vidyapeetham, India}

\author{Manu Kurian}
\email{manu.kurian@iitgn.ac.in}
\affiliation{Indian Institute of Technology Gandhinagar, Gandhinagar-382355, Gujarat, India \\ 
Department of Physics, McGill University, 3600 University Street, Montreal, QC, H3A 2T8, Canada}

\author{Vinod Chandra}
\email{vchandra@iitgn.ac.in}
\affiliation{Indian Institute of Technology Gandhinagar, Gandhinagar-382355, Gujarat, India}


\begin{abstract}

The effects of collisional processes in the hot QCD medium to thermal dilepton production from 
$q\overline{q}$ annihilation in relativistic heavy-ion collisions have been investigated. The 
non-equilibrium corrections to the momentum distribution function have been estimated within 
the framework of ensemble-averaged diffusive Vlasov-Boltzmann equation, encoding the effects 
of collisional processes and turbulent chromo-fields in the medium. 
The contributions from the $2\rightarrow2$ elastic scattering processes 
have been quantified for the thermal dilepton production rate. It is seen that the collisional corrections enhance 
the equilibrium dilepton spectra at high $p_T$ and suppress at lower $p_T$.
A comparative study between collisional and anomalous contributions to the 
dilepton production rates has also been explored. The collisional contributions 
are seen to be marginal over that due to collisionless anomalous transport.

\end{abstract}

\maketitle

\section{Introduction}

Heavy-ion collision experiments at Relativistic Heavy Ion Collider (RHIC), 
and at Large Hadron Collider (LHC), CERN enable the creation of the strongly 
coupled matter quark-gluon plasma (QGP)~\cite{expt}, which is assumed to have existed in the very 
early universe a few microseconds after Big-Bang~\cite{Hatsuda}. 
In these experiments, heavy nuclei are collided at 
ultra-relativistic energies to produce an expanding hot fireball. The created matter expands in 
space and time followed by hadronization with the decrease of temperature (with a cross-over from the deconfined quarks and gluons to hadrons for most of the heavy-ion collisions).
The evolution of QGP is successfully studied 
within the framework of causal relativistic hydrodynamics~
\cite{Heinz,Muronga:2003ta,Romatschke:2007mq,Gale:2013da,Romatschke:2017ejr}. 
These studies, along with the experimental observations, suggest the existence of strongly coupled QGP 
with near-perfect fluid nature~\cite{hirano}. {The momentum asymmetry of the QCD medium present during the 
entire medium expansion may induce instabilities to the chromo-field equations. These instabilities
in the rapidly expanding medium can lead to the plasma 
turbulence in the heavy-ion collisions~\cite{Dupree,Asakawa:2006tc}. 
These give rise to the anomalous 
transport in the medium. The impacts of anomalous transport coefficients due to the momentum asymmetry in the medium are 
well investigated in the electromagnetic plasmas~\cite{Abe,Okada} and 
hot QCD plasmas~\cite{Asakawa:2006tc,Majumder:2007zh,Asakawa:2006jn}. There have been several studies on the 
collisional contributions to the transport parameters for the hot QCD medium~\cite{Mitra:2016zdw,Kurian:2019fty}.
The interplay of anomalous and collisional corrections can be studied by analyzing the signals emitted 
from various phases of fireball expansion. 

Thermal dileptons and photons are one 
of the most efficient probes of QGP~
\cite{Feinberg76,Shyryak78,McLerran:1984ay,Kajantie:1986dh,Rapp:2014hha}. 
Since they interact electromagnetically, these thermal radiations can reach the detectors without 
being rescattered. These radiations are emitted throughout the expansion of the 
fireball with negligible final-state interactions and can carry information about the hotter phases of the matter as well as the initial state of QGP after collision~
\cite{Alam:1996fd,Alam:1999sc,Peitzmann:2001mz}. 
The dilepton invariant mass spectrum has contributions from various processes throughout 
the evolution of fireball. The high mass range ($M>3$ GeV) has contributions from the hadronic reactions such as photoproduction processes and jet-dilepton conversion arising from initial hadronic scattering. The decays of vector mesons have a considerable contribution in 
the low mass range $0.6<M<1.1$ GeV. For $M<0.2$ GeV, we encounter the contribution of pion 
decays from the hadronic phase. While thermal dileptons from QGP are prominent in the 
intermediate-mass range, $0.2<M<2.5$ GeV with major contribution coming from the 
$q\bar{q}$ annihilation process~\cite{Rapp:1999ej,Alam:1999sc,Vujanovic:2013jpa}. 

{The anisotropic and viscous effects on dilepton production have been investigated, 
such as dissipative effects due to shear viscosity~\cite{Dusling:2008xj,Bhatt:2011kx,Bhalerao:2013aha,Chandra:2015rdz,
Vujanovic:2017psb}.} The correction due to bulk viscosity on dilepton 
production was introduced~\cite{Bhatt:2011kx} and studied 
too~\cite{Bhalerao:2013aha,Chandra:2015rdz,Vujanovic:2019yih}.  
Recently some works have been done to understand the effect of vorticity and magnetic field 
in the thermal dilepton production
~\cite{Bhatt:2018xsx,Das:2019nzv,Bandyopadhyay:2016fyd,Ghosh:2018xhh}. 
In Ref.~\cite{Chandra:2016dwy}, the effect of chromo-Weibel instability in the dilepton 
production rate has been investigated. It will be an interesting task to study the 
collisional contributions to thermal dilepton production along with the anomalous corrections. 
The first step towards this analysis is the proper modelling of the near-equilibrium momentum 
distribution functions for quarks and gluons while incorporating the effects of 
collisional aspects and anomalous 
transport in the medium. This sets the motivation for the present analysis.   

The static dilepton production rate from $q\bar{q}$ annihilation 
in the presence of these corrections is obtained from the relativistic kinetic theory. 
The dilepton rate is calculated by incorporating the QCD medium interaction effects
in the cross-section. The total 
dilepton yield depends on the temperature profile of the expanding QGP. This is obtained 
from hydrodynamic modelling by providing appropriate initial conditions and 
realistic equation of state (EoS). 
It is crucial to note that the role of EoS is important in analyzing the signals from QGP, 
such as thermal photons~\cite{Bhatt:2010} and dileptons~\cite{Bhatt:2011kx}. 
Here, we employ an effective fugacity quasi-particle model 
(EQPM)~\cite{Chandra:2011en,Chandra:2007ca} to incorporate the 
realistic EoS effects in the analysis. 
In the current analysis, the near-equilibrium distribution functions are obtained as the modification over these distributions induced by anomalous and collisional processes within an effective transport approach closely following Refs.~\cite{Asakawa:2006jn,Chandra:2008hi}. 
The distribution functions thus obtained have been employed to study thermal dilepton spectra. 

The paper is organized as follows.  Section II is devoted to the estimation of 
non-equilibrium phase-space 
momentum distribution function while incorporating the effects of collisional processes, 
and anomalous 
transport in the QGP medium is described. In Section III, the thermal dilepton production rates are computed 
in the presence of collisional processes along with the turbulent fields. Section IV deals with dilepton 
yields for an expanding QGP in heavy-ion collisions. The results and followed discussions are presented 
in section V, and finally, we conclude the analysis with an outlook in section VI.

{\bf Notations and conventions}:  We are working in units with $k_{B}=1$, $c=1$, $\hbar=1$. {The signature of Minkowski metric used is $\eta_{\mu\nu}= \textrm{diag}(+,-,-,-)$}.
The term $u^{\mu}$ denotes 
the fluid four-velocity and is normalized to unity $u^{\mu}u_{\mu}=1$. In the fluid rest frame, $u^{\mu}=(1, 0, 0 ,0)$.
The quantity $\Delta u_{\mu\nu}=\frac{1}{2}(\nabla_\mu u_\nu+\nabla_\nu u_\mu)-\frac{1}{3}\Delta_{\mu\nu}\nabla_\gamma
u^\gamma$ defines the traceless symmetric velocity gradient, with $\Delta_{\mu\nu} \equiv \eta_{\mu\nu} - u_\mu u_\nu$ being the projection operator orthogonal to $u^\mu$ and 
$\nabla_\mu = \Delta_{\mu\nu}\partial^\nu$.

\section{Modified quark (antiquark) distribution function}

Here, a momentum anisotropic hot QCD medium is considered, keeping the fact in mind that momentum 
anisotropy may sustain in the latter stage of the collisions. The momentum anisotropy may lead to 
chromo-Weibel instability whose physics is captured in terms of an effective diffusive Vlasov-Boltzmann term depicting the anomalous transport in the 
hot QCD medium~\cite{Asakawa:2006tc}. It has been argued that the anomalous viscosity dominates over the collisional viscosity in the regime of weak coupling~\cite{Asakawa:2006jn}.
The linear transport equation in the presence of turbulent color fields with a collisional term where $2\rightarrow2$
elastic contributions have been taken into account. The ansatz for the momentum anisotropic/near-equilibrium distribution
functions for gluonic and quark/anti-quark degrees of freedom in hot QCD medium is of the form,
\ba\label{1}
f_{g/q}({\vec{p}},{\vec{r}}) = f_{0~g/q}\left[1+(1 \pm f_{0~g/q})f_{1~g/q}({\vec{p}})\right],
\ea
where $f_{0}$ and $f_1$ denotes the equilibrium and linear order perturbation to the distribution function, respectively.
Before obtaining the deviation of momentum distribution function away from equilibrium that 
encodes the effects of anomalous transport as well as collisional processes, adequate modelling of 
equilibrium distribution functions for the gluons and quarks has to be considered to incorporate the 
realistic EoS in the analysis. 
The EQPM employed in the current analysis interprets the thermal QCD 
medium EoS with the non-interacting quasigluons and quasiquarks/antiquarks with effective fugacities $z_{g/q}$
and have the following distribution functions~\cite{Chandra:2011en,Chandra:2007ca},
\begin{equation}
f_{0~g/q}=\frac{z_{g/q}\exp{(-\beta E_{p})}}{1\mp z_{g/q}\exp{(-\beta E_{p})}},
\end{equation}    
where $E_{p}=\mid \vec{p}\mid \equiv p$ for gluons and  $E_{p}=\sqrt{p^{2}+m^{2}}$ for quarks. The physical significance of the effective fugacity parameter can be understood from the non-trivial energy dispersion relation,
\begin{align}
\omega_{p} = E_p + \delta \omega_{g/q}, &&\delta \omega_{g/q} = T^{2}\partial_{T} \ln(z_{g/q}),
\end{align}
where $\delta \omega_{g/q}$  
denotes the modified part of the dispersion relation. The temperature dependence 
 of effective fugacities can be obtained from the lattice EoS~\cite{Chandra:2011en}.
The fugacity parameters are not associated with any conserved 
current in the medium and retain the same form for quarks and antiquarks~\cite{Bhadury:2019xdf}. 
%
The QCD thermodynamics has been studied within the EQPM description, and the results have been compared with that of the lattice data~\cite{Chandra:2011en}. It is observed that the EQPM results are in agreement with the lattice data beyond the transition temperature. In particular, the model accurately describes the trace anomaly of the medium. Further, an effective covariant kinetic theory has been developed within the EQPM to study the near-equilibrium dynamics of the QCD medium~\cite{Mitra:2018akk}.  It is important to emphasize that a Virial expansion for the QCD medium has been obtained in terms of quasiparticle number densities to study the QCD interaction. The comparison of the EQPM with other approaches (effective mass model, models based on Polyakov loop, etc.) has been conducted in Ref.~\cite{Chandra:2011en}. The EQPM and the followed effective transport theory approach have been employed to study the transport properties~\cite{Mitra:2017sjo,Kurian:2018qwb} and momentum anisotropy~\cite{Kumar:2017bja,Jamal:2018mog} of the QCD medium.

We choose the following ansatz for the linear order perturbation $f_{1~g/q}({\vec{p}})$ to the isotropic gluon and quarks
distribution functions 
\begin{equation}
  f_{1~g/q}({\vec{p}}) = \frac{\bar{\Delta}_{1~g/q}({\vec{p}})}{\omega_{p} T^2 } p^\mu p^\nu \Delta u_{\mu\nu}.
\end{equation}
In the local rest frame of the fluid, we can write $\Delta u_{ij} = -[\frac{1}{2}(\nabla_i u_j + \nabla_j u_i)-\frac{1}{3}\delta_{ij} \nabla \cdot u]$.
Now, considering the boost invariant longitudinal Bjorken's flow, with $u=\frac{z}{\tau}$ and 
$\Delta u_{ij}=-\frac{1}{3\tau}\text{diag}(-1, -1, 2)$, the expression for $f_{1~g/q}({\vec{p}})$ becomes
\ba
f_{1~g/q}({\vec{p}}) =- \frac{\bar{\Delta}_{1~g/q}({\vec{p}})}{\omega_{p} T^2 \tau}\left(p_{z}^2 - \frac{p^2}{3} \right),
\label{eq:f1g1q}
\ea
where $\tau$ is the proper time.
In the current analysis, we investigate two dominant sources of the non-equilibrium dynamics of the QGP medium, namely, momentum anisotropy of the QGP medium, and collisional processes in the medium. The strength of the turbulent fields depends on the magnitude of the momentum anisotropy in the medium and gives rise to anomalous transport in the medium. Both these non-equilibrium effects will contribute to shear viscosity as the relaxation rates due to both processes are additive (anomalous and collisional contributions to the viscosity as described in Ref.~\cite{Asakawa:2006tc}). These non-equilibrium effects are embedded in the analysis through $f_{1~g/q}({\vec{p}})$. The momentum anisotropy may induce instabilities in the rapidly expanding medium, which can lead to plasma turbulence, and the turbulent fields give rise to anomalous transport in the medium. The non-equilibrium corrections to the momentum distribution function can be estimated within the framework of
ensemble-averaged diffusive Vlasov-Boltzmann equation, encoding the effects of collisional processes
and turbulent chromo-fields in the medium. The quantity $\bar{\Delta}_{1~g/q}({\vec{p}})$ captures the strength of non-equilibrium part of the distribution function.  For the case with only anomalous transport (where collisional aspects are negligible) $\bar{\Delta}_{1~g/q}({\vec{p}})$ can be defined as follows~\cite{Asakawa:2006tc, Chandra:2016dwy}, 
\begin{equation}
\bar{\Delta}_{1~g/q}({{\vec{p}}})= 2(N_c^2-1)\dfrac{\omega_{g/q}T}{3g^2C_2\langle E^2+B^2\rangle_{g/q}\tau_m},
\end{equation} 
where unknown factors in the denominator is related to the jet quenching parameter $\hat{q}$. Here, $\langle E^2\rangle$ and $\langle B^2\rangle$ represent the color averaged chromo-electromagnetic fields and $\tau_m$ measures the time scale of instability in the medium. The current focus is to 
incorporate the collisional aspects along with the anomalous contributions to the momentum distribution functions. For the general case, $\bar{\Delta}_{1~g/q}({\vec{p}})$ takes the following form,
\ba   
\bar{\Delta}_{1~g/q}({\vec{p}}) = \frac{{A_{g/q}^t}|\vec{p}|}{T}.
\label{eq:Delta}
\ea

{Here, the quantity $A_{g/q}^t$ has contributions from both anomalous and collisional transports.}
The effect of collisional processes 
in the evolution of distribution function can be quantified with the collision 
kernel in the transport equation.  
From Eq.~(\ref{eq:f1g1q}), the leading order correction to the quark 
distribution function can be considered as,
\begin{equation} \label{deltaf_1}
  f_{1~q}=-\frac{{A_{q}^t}|\vec{p}|}{\omega_{p}(T,\vec{p}) T^3 \tau}\left(p_z^2-\frac{p^2}{3}\right),
\end{equation}
where $\omega_{p}^{-1}$ takes the following form in the linear expansion,
\begin{equation}\label{omega1}
\omega_{p}^{-1}(T,\vec{p})\approx\left[\frac{1}{|\vec{p}|}-
\frac{\delta \omega_q}{|\vec{p}|^2}\right].
\end{equation}
The authors of the Ref.~\cite{Mitra:2017sjo} have realized that the leading order (in temperature gradient of effective fugacity) correction to single particle energy is less than $10\%$ at $T=2.5T_c$ and observed considerable agreement in the results 
of transport coefficients from full numerical coding and from the linear expansion approximation.}
Note that we are not considering the subscript for quarks while defining the distribution function and dispersion relation as the current focus is on the dilepton production by $q\bar{q}$ annihilation. Combining Eqs.~(\ref{deltaf_1}) and (\ref{omega1}), we obtain the form of $f_{1~q}$ as
\begin{eqnarray}\label{deltaf}
 f_{1~q}=-\frac{{A_{q}^t}}{T^3\tau}\left[1-\frac{\delta \omega_q}
 {|\vec{p}|}\right]\left(p_{z}^2 - \frac{p^2}{3} \right).
\end{eqnarray}
Similarly, one can estimate the non-equilibrium corrections to the 
gluon distribution function in terms of ${A_{g}^t}$.

\subsubsection*{The form of ${A_{g/q}^t}$}

Following the same formalism as in Refs.~\cite{Chandra:2016dwy, Asakawa:2006jn},  one can estimate the algebraic equation of ${A_{g/q}^t}$ by taking the appropriate moment of the transport equation. The ensemble average Vlasov-Boltzmann equation takes the form as follows,
\begin{equation}\label{R1.1}
v^{\mu}\dfrac{\partial}{\partial x^{\mu}}\bar{f}-\mathcal{F}_A\bar{f}+\langle C[f]\rangle=0,
\end{equation}
where $\bar{f}$ is the ensemble-averaged thermal distribution of the particles. In our case, $\bar{f}\equiv f_{g/q}$ as defined in the Eq.~(\ref{1}). The diffusive Vlasov term characterizes the contribution from turbulent fields, 
and the force term takes the form as,
\begin{align}\label{R1.2}
\mathcal{F}_A\bar{f}=&\dfrac{g^2C_f}{3(N_c^2-1)\omega_{g, q}}\langle E^2+B^2\rangle_{g, q}\tau_m\nonumber\\
&\times \mathcal{L}^2f_{0~g/q}(1\pm f_{0~g/q})p_ip_j\Delta u_{ij},
\end{align}
with $C_f$ is the Casimir invariant of $SU(N_c)$ gauge theory and the 
operator $ \mathcal{L}^2$ can be defined as,
\begin{equation}
 \mathcal{L}^2=\mid\vec{p}\times\partial_{\vec{p}}\mid^2-\mid\vec{p}\times\partial_{\vec{p}}\mid^2_z.
\end{equation} 
{In the current analysis, $A^t$ has contributions from both anomalous and collisional transports. 
For the anomalous transport, the strength of the turbulent fields are related to the anomalous transport in the medium. In addition, we have switched on the collisional term in the Boltzmann equation which contribute to the collisional aspects of the medium.}
The collision kernel $\langle C[f]\rangle$ in the transport equation measures the leading order contributions from the collisional processes. The collision 
integral for the $2\rightarrow2$ scattering 
process $\vec{p}, \vec{k}\rightarrow \vec{p}^{'}, \vec{k}^{'}$ is defined as~\cite{Asakawa:2006jn,Arnold:2000dr},
\begin{align}
C[f]=&\dfrac{1}{4E_p}\int{\dfrac{d^3{\vec{k}}}{(2\pi)^32E_k}}\int{\dfrac{d^3 {\vec{p}}^{'}}{(2\pi)^32E_{p^{'}}}}\int{\dfrac{d^3{\vec{k}}^{'}}{(2\pi)^32E_{k^{'}}}}\nonumber\\
&\times|\mathcal{M}|^2(2\pi)^4\delta^4(P+K-P^{'}-K^{'})\nonumber\\
&\times \Bigg[f_{g/q}({\vec{p}})f_{g/q}({\vec{k}}) \Big(1\pm f_{g/q}({\vec{p}}^{'})\Big) \Big(1\pm f_{g/q}({\vec{k}}^{'})\Big)\nonumber\\
&-f_{g/q}({\vec{p}^{'}})f_{g/q}({\vec{k}})\Big(1\pm f_{g/q}({\vec{p}})\Big) \Big(1\pm f_{g/q}({\vec{k}})\Big)\Bigg],
\end{align}
where $|\mathcal{M}|^2$ is the scattering amplitude and $P, K, P^{'}$ and $K^{'}$ are 
the four-momenta of the particles before and after scattering. The linearized transport equation is a linear 
integral equation, and one can employ variational method by minimizing the linearized Vlasov-Boltzmann 
equation to determine $\bar{\Delta}_{1~g/q}({\vec{p}}) $. Following this standard method as in 
Refs.~\cite{Chandra:2016dwy, Asakawa:2006jn} leads to the following 
matrix equation for the column vector ${A^t =\{{A_{g}^t},{A_{q}^t}\}}$,
\ba
(\tilde{a}_A+\tilde{a}_C){A^t} = \tilde{r}.
\label{eq:A}
\ea
The column vector $\tilde{r}$ takes the form,
\ba 
\tilde{r} = \left\{\frac{32 \left(N_c^2-1\right) I_5^g}{3 \pi ^2},\frac{32 N_c N_f I_5^q}{3 \pi ^2}\right\},
\label{eq:r}
\ea
where $N_f$ is the number of flavors, and the function $I_n$ takes the form,
\begin{align}
 &I_n^{q}=-\rm{PolyLog}[n, -z_{q}],  &&I_n^{g}=\rm{PolyLog}[n, z_{g}],
\end{align}
for quarks and gluons. 
The matrices $\tilde{a}_A$ and $\tilde{a}_C$ denote the anomalous transport and 
collisional (elastic scattering processes) contribution of the transport equation. 
The matrix $\tilde{a}_A$ takes the following form,
\ba
\tilde{a}_{A} = \left(
\begin{array}{cc}
 \frac{32 N_c I_4^g Q_g}{5 \pi ^2 T^3} & 0 \\
 0 & \frac{32 N_f I_4^q Q_q}{5 \pi ^2 T^3} \\
\end{array}
\right),
\label{eq:cA}
\ea
where $Q_{g/q}$ is defined as,
\ba
Q_{g/q}=\frac{g^2\langle E^2+B^2\rangle_{g, q}}{2} \tau_m.
\ea
The matrix $\tilde{a}_C$ can be described as follows,
\begin{widetext}
\ba
 \tilde{a}_{C} =C_{c} \left(
\begin{array}{cc}
 \frac{7  N_c \left(2 N_c+N_f\right) I_2^g}{24 \pi ^2 z_g}+\frac{ N_c \left(N_c^2-1\right) N_f \left(I_4^g+I_4^q\right) z_g}{2 \pi ^3 \left(z_g+z_q\right)} & -\frac{ N_c \left(N_c^2-1\right) N_f \left(I_4^g+I_4^q\right) z_g}{2 \pi ^3 \left(z_g+z_q\right)} \\
 -\frac{ N_c \left(N_c^2-1\right) N_f \left(I_4^g+I_4^q\right) z_g}{2 \pi ^3 \left(z_g+z_q\right)} & \frac{7 N_f \left(2 N_c+N_f\right) I_2^q}{24 \pi ^2 z_q}+\frac{ N_c \left(N_c^2-1\right) N_f \left(I_4^g+I_4^q\right) z_g}{2 \pi ^3 \left(z_g+z_q\right)} \\
\end{array}
\right).
\label{eq:cC}
\ea
\end{widetext}
The EQPM is based on the charge 
renormalization in medium~\cite{Chandra:2011en}, and one can define an effective coupling 
$\alpha_{eff}$ by investigating the Debye 
screening mass of the QCD medium~\cite{Mitra:2017sjo}. Note that in the 
leading-log order, we have 
$C_{c}\approx 2\pi^2(N_c^2 - 1)\alpha^2_{eff}\log(\alpha_{eff}^{-1})$, where $\alpha_{eff}$ takes the form,
\begin{equation}
\alpha_{eff}=\alpha_s (T)\Bigg(\dfrac{\frac{2N_c}{\pi^2}I_2^g+\frac{2N_f}{\pi^2}I^q_2}{\frac{N_c}{3}+\frac{N_f}{6}}\Bigg).
\end{equation}
The 2-loop expression for QCD running coupling constant $\alpha_s(T)$ at 
finite temperature can be defined as~\cite{Srivastava:2015via,Haque:2012my,Laine:2005ai}, 
\be
\alpha_s (T)=\dfrac{6\pi}{(33-2N_f)\ln{\frac{T}{\Lambda_T}}}
\Bigg(1-\dfrac{3(153-19N_f)}{(33-2N_f)^2}\dfrac{\ln(2\ln{\frac{T}{\Lambda_T}})}{\ln{\frac{T}{\Lambda_T}}}\Bigg),
\ee
with QCD scale fixing parameter can be defined from the $\overline{MS}$ scheme such that $\Lambda_T=\frac{\exp(\gamma_E+1/22)}{4\pi}\Lambda_{MS}$, where $\gamma_E=0.5772156$ and the renormalization scale $\Lambda_{MS}=1.14~T_c$~\cite{Chandra:2007ca}. 
The algebraic forms of ${A_{g}^t}$ and ${A_{q}^t}$ can be obtained by solving Eq.~(\ref{eq:A}) using Eqs.~(\ref{eq:r}),~(\ref{eq:cA}) and ~(\ref{eq:cC}) and have the following forms,
\ba
{{A_{g}^t}}= \frac{1280 \pi  \left(N_c^2-1\right)\Big(N_c^2N_fI^q_5~\phi_0+I_5^g ~ \phi_1\Big)}{N_c\Big(-N_c(N_c^2-1)^2N_f\phi^2_0+\phi_1\phi_2\Big)},
\ea
and
\begin{equation}
 {{A_{q}^t}}=\frac{1280\pi \Big((N_c^2-1)^2I^g_5~\phi_0+N_c I_5^q~\phi_2\Big)}{-N_c(N_c^2-1)^2N_f\phi^2_0+\phi_1 \phi_2}.
\end{equation}
The quantities $\phi_0$, $ \phi_1$ and $ \phi_2$ are defined as follows,
\begin{eqnarray}
\phi_0&=&60 C_c \Big(I_4^g+I_4^q\Big)\frac{ z_g }{z_g+z_q},\nonumber\\
 \phi_1&=&N_c(N_c^2-1)\phi_0+\frac{35\pi C_c  I_2^q(2N_c+N_f)}{z_q}+\frac{768\pi Q_q I_4^q}{T^3},\nonumber \\
 \phi_2&=& (N_c^2-1)N_f \phi_0+\frac{35\pi C_c I_2^g(2 N_c+N_f)}{z_g}+\frac{768\pi Q_g I_4^g}{T^3}\nonumber.
\end{eqnarray}
{
It is important to emphasize that the expressions of ${A_{g}^t}$ and ${A_{q}^t}$ reduce to the 
results of Ref.~\cite{Asakawa:2006jn,Chandra:2008hi} in the limit $z_{g,q} \rightarrow 1$ and in the absence of collisions, $ \tilde{a}_{C}=0$. }

Next, by employing these non-equilibrium distribution functions, we study the thermal dilepton production from 
hot QCD medium. 

\section{Thermal dilepton production rate}

Thermal dileptons produced in the QGP medium has major contributions from the $q\bar{q}$ annihilation process, 
$q\bar{q}\rightarrow\gamma^*\rightarrow l^+ l^-$. The rate of dilepton production for this process within
the EQPM model in terms of 
quark distribution function can be defined as,
\begin{eqnarray} \label{rate_m}
 \frac{dN}{d^4x d^4p} &=& \int \frac{d^3\vec{p}_1}{(2\pi)^3} \frac{d^3\vec{p}_2}{(2\pi)^3}\,
 \frac{M_{eff}^2 g_d^2 \sigma(M_{eff}^2)}{2 \omega_1 \omega_2}\nonumber\\
 && \times f(\vec{p}_1) f(\vec{p}_2) \delta^4(\tilde{p}-\tilde{p}_1-\tilde{p}_2).
\end{eqnarray}
Note that the subscript $q$ for the distribution function is dropped from this section as the focus is only 
on the $q\bar{q}$ annihilation process, $i.e.$, $f_q(\vec{p})\equiv f(\vec{p})$ 
and is described in Eq.~(\ref{1}).
Here, the quantity $M_{eff}$ is the medium modified effective mass 
of the virtual photon in the interacting QCD medium with
\begin{eqnarray}
M_{eff}^2&=(\omega_1 + \omega_2)^2-(\vec{p}_1 + \vec{p}_2)^2\nonumber \\&\approx 
M^2 \left( 1 + \frac{4\delta \omega_q (E_1 + E_2)}{M^2}\right),
\end{eqnarray}
where $M^2$ represents its invariant mass in the limit of $z_{q/g}=1$. 
The quantity $\tilde{p}_{1,2}=(\omega_{1,2},\,\vec{p}_{1,2})$ is the 
4-momenta of the quark and antiquark respectively and $\tilde{p}=(\omega_0=\omega_1+\omega_2,\,
\vec{p}=\vec{p}_1+\vec{p}_2)$ 
is the 4-momentum of the dilepton pair. If the 
quark masses are neglected, we can write $\omega_{1,2}=\sqrt{{\vec{p}_{1,2}}^2+m^2}\approx|\vec{p}_{1,2}|$. 
The term $\sigma(M_{eff}^2)$ is the thermal dilepton production cross section and $g_d$ is the degeneracy factor.
The relative velocity of the quark-antiquark pair is given by 
$v_{rel}=\sqrt{\frac{M_{eff}^2(M_{eff}^2-4m^2)}{4\omega_1^2\omega_2^2}}\simeq
\frac{M_{eff}^2}{2 \omega_1 \omega_2}$.
With $N_f=2$ and $N_c=3$, we have $M_{eff}^2g_d^2\sigma(M_{eff}^2)=\frac{80\pi}{9}\alpha^2$. 
We are interested in the regime in which invariant masses are larger than the temperature, $M>>T>>m$. 
Hence we can approximate the Fermi-Dirac distribution by that of the classical Maxwell-Boltzmann distribution (high temperature limit)
i.e., $f_0(\vec{p})\backsimeq z_q e^{-\omega/T}$. 
Under this approximation, the quark (antiquark) distribution function, 
described by Eq.~(\ref{1}) and Eq.~(\ref{deltaf}), in the covariant form, becomes:  
\begin{eqnarray}
 f(\vec{p}) \backsimeq z_q e^{-\omega/T} \left[1 + {\chi^t(\vec{p}, T)}
 \tilde{p}^\mu \tilde{p}^\nu \Delta u_{\mu\nu} \right];
\end{eqnarray}
{where quasiparticle four-momenta ($\Tilde{p}^{\mu}$) are related to the bare momenta ($p^{\mu}$) as $\tilde{p}^{\mu} = p^{\mu}+\delta\omega_{g/q}\, u^{\mu}$} and the quantity ${\chi^t(\vec{p}, T)}$ represents,
\begin{equation}\label{chi}
{ \chi^t(\vec{p}, T)}
 = \frac{{A_q^t}}{T^3} \left[1 - \frac{\delta \omega_q}{|\vec{p}|} \right].
\end{equation}
Keeping the terms only upto quadratic order in momenta, the dilepton production rate takes the form as follows,
\begin{eqnarray}
\frac{dN}{d^4x d^4p} &=& \int \frac{d^3 \vec{p}_1}{(2\pi)^3} \frac{d^3 \vec{p}_2}{(2\pi)^3}
\frac{M_{eff}^2 g_d^2 \sigma(M_{eff}^2)}{2 \omega_1 \omega_2} \nonumber \\
&& \times \left[1 + 2\,{\chi^t(\vec{p}, T)}\,\tilde{p}_1^\mu \tilde{p}_1^\nu \Delta u_{\mu\nu}\right]\nonumber \\
&& \times f_0(\vec{p}_1) f_0(\vec{p}_2)\,\delta^4(\tilde{p}-\tilde{p}_1-\tilde{p}_2)\nonumber\\
&=& \frac{dN_0}{d^4x d^4p} + \frac{dN_\chi}{d^4x d^4p}; \label{drate_tot}
\end{eqnarray}
where the equilibrium contribution to dilepton production takes the form
\begin{eqnarray}
 \frac{dN_0}{d^4x d^4p} &=& \int \frac{d^3 \vec{p}_1}{(2\pi)^6}\,
 \frac{M_{eff}^2 g_d^2 \sigma(M_{eff}^2)}{2 \omega_1 \omega_2} z_q^2 e^{-(\omega_1 + \omega_2)/T} \nonumber\\
 && \times \delta(\omega_0 -\omega_1 -\omega_2) \nonumber \\
 &=& \frac{z_q^2}{2} \frac{M_{eff}^2 g_d^2 \sigma(M_{eff}^2)}{(2\pi)^5}e^{-\omega_0/T}.
\end{eqnarray}
Using Eq.~(\ref{omega1}),~(\ref{chi}) in~(\ref{drate_tot}), we write the non-equilibrium contribution to the dilepton production rate as,
\begin{eqnarray}
 \frac{dN_\chi}{d^4x d^4p} &=& 2  \int \frac{d^3 \vec{p}_1}{(2\pi)^6}\,
 \frac{M_{eff}^2 g_d^2 \sigma(M_{eff}^2)}{2 \omega_1 \omega_2}
 z_q^2 e^{-(\omega_1 + \omega_2)/T}  \nonumber\\
 &&  \times \frac{A_q^t}{T^3} \left[1 -\frac{\delta \omega_q}{|\vec{p}_1|}\right]\,
 \tilde{p}_1^\mu \tilde{p}_1^\nu \Delta u_{\mu\nu}  \nonumber\\
 &&  \times \delta(\omega_0 -\omega_1 -\omega_2)  \nonumber\\
&=& I^{\mu\nu}(p) \Delta u_{\mu\nu};
\end{eqnarray}
where, 
\begin{eqnarray}\label{I.1}
 I^{\mu\nu}(p) &=& 2  \int \frac{d^3 \vec{p}_1}{(2\pi)^6}\,
 \frac{M_{eff}^2 g_d^2 \sigma(M_{eff}^2)}{2 \omega_1 \omega_2}z_q^2 e^{-(\omega_1 + \omega_2)/T}  \nonumber\\
 &&  \times \frac{A_q^t}{T^3} \left[1 -\frac{\delta \omega_q}{|\vec{p}_1|}\right]\,
 \tilde{p}_{1}^\mu \tilde{p}_{1}^\nu \delta(\omega_0 -\omega_1 -\omega_2) .
\end{eqnarray}
The general form of the second rank tensor can be 
constructed using the metric $\eta^{\mu\nu}$, $u^\mu$ and $\tilde{p}^\mu$ as,
\begin{eqnarray}
 I^{\mu\nu}(p)=a_0 \eta^{\mu\nu} + a_1 u^\mu u^\nu + a_2 \tilde{p}^\mu \tilde{p}^\nu
 + a_3(u^\mu \tilde{p}^\nu + u^\nu \tilde{p}^\mu). \label{I_tensor} \nonumber\\
\end{eqnarray}
Since $u^\mu\Delta u_{\mu\nu}=0$ and $\eta^{\mu\nu} \Delta u_{\mu\nu}=0$, only the 
coefficient $a_2$ survives when Eq.~($\ref{I_tensor}$) is 
contracted with $\Delta u_{\mu\nu}$. Moreover, we construct 
a projection operator $P_{\mu\nu}$ such that $a_2 = P_{\mu\nu}I^{\mu\nu}$. 
The form of $P_{\mu\nu}$ can be obtained as
\begin{eqnarray}\label{P.1}
 P_{\mu\nu} &=& \frac{1}{2|\vec{p}|^4} \Big[|\vec{p}|^2\eta_{\mu\nu} + (2\omega_0^2 + M_{eff}^2)u_\mu u_\nu
 +3 \tilde{p}_\mu \tilde{p}_\nu \nonumber\\
 &&- 6 \omega_0 u_\mu \tilde{p}_\nu \Big].
\end{eqnarray}
Incorporating these steps, the non-equilibrium contribution to dilepton rate takes the form as follows,
\begin{eqnarray}
 \frac{dN_\chi}{d^4x d^4p} = a_2\tilde{p}^\mu \tilde{p}^\nu  \Delta u_{\mu\nu}
 = \left \lbrace P_{\alpha\beta} I^{\alpha\beta} \right \rbrace
 \tilde{p}^\mu \tilde{p}^\nu \Delta u_{\mu\nu}.\nonumber\\
\end{eqnarray}
Employing Eq.~(\ref{I.1}) and Eq.~(\ref{P.1}) we have,
\begin{eqnarray}
 P_{\alpha\beta}I^{\alpha\beta} &=& \frac{ A_q^t}{|\vec{p}|^4T^3} \int \frac{d^3 \vec{p}_1}{(2\pi)^6}\,
 \frac{M_{eff}^2 g_d^2 \sigma(M_{eff}^2)}{2 \omega_1 \omega_2}z_q^2 e^{-(\omega_1 + \omega_2)/T} \nonumber\\
 && \times\left[(2\omega_0^2 + M_{eff}^2)\omega_1^2 + 
 3 (\tilde{p}\cdot \tilde{p}_1)^2 
 -6\omega_0 \omega_1(\tilde{p} \cdot \tilde{p}_1)\right] \nonumber\\
 &&\times \left[1-\frac{\delta \omega_q}{|\vec{p}_1|}\right] \delta(\omega_0-\omega_1-\omega_2)\nonumber\\
 &=& \frac{A_q^t}{T^3} \frac{1}{2 |\vec{p}|^5} [\mathscr{M} + \mathscr{N}],
\end{eqnarray}
where $ \mathscr{M}$ and $ \mathscr{N}$ can be defined as,
\begin{eqnarray}
\mathscr{M} &=& \int dp_1 \frac{M_{eff}^2 g_d^2 \sigma(M_{eff}^2)}{(2\pi)^5} z_q^2 \,e^{-\omega_0/T}\nonumber\\
&& \times \left[(3\omega_0^2 - |\vec{p}|^2) |\vec{p}_1|^2 - 3\omega_0 M_{eff}^2 |\vec{p}_1|
+ \frac{3}{4}M_{eff}^4 \right] \nonumber\\
&=& \frac{dN_0}{d^4x d^4p} \frac{4}{3}|\vec{p}|^5, \nonumber\\
\textrm{and}  \nonumber\\
\mathscr{N} &=& -\delta \omega_q\int dp_1 \frac{M_{eff}^2 g_d^2 \sigma(M_{eff}^2)}{(2\pi)^5} z_q^2\,
e^{-\omega_0/T}  \nonumber\\
 && \times \left[(3\omega_0^2 - |\vec{p}|^2) |\vec{p}_1| - 3\omega_0 M_{eff}^2 
 + \frac{3}{4}\frac{M_{eff}^4}{|\vec{p}_1|}  \right] \nonumber\\
&=& -\frac{dN_0}{d^4x d^4p} 2\delta \omega_q  \Bigg[(2\omega_0^2-5M_{eff}^2)\frac{\omega_0|\vec{p}|}{2}  
\nonumber\\
&& +\frac{3}{4} M_{eff}^4 \ln \left(\frac{\omega_0 +|\vec{p}|}{\omega_0-|\vec{p}|}\right)\Bigg] \nonumber
\end{eqnarray}
respectively. Thus, we obtain the non-equilibrium contribution to the 
dilepton rate as,
\begin{eqnarray}
 \frac{dN_\chi}{d^4x d^4p}&=&\frac{dN_0}{d^4x d^4p}\frac{A_q^t}{T^3}
 \Bigg\lbrace \frac{\delta \omega_q}{|\vec{p}|^5}\Big[(5M_{eff}^2 - 2\omega_0^2)
 \frac{\omega_0|\vec{p}|}{2} \nonumber\\
&&-\frac{3}{4}M_{eff}^4\ln \left(\frac{\omega_0+|\vec{p}|}{\omega_0-|\vec{p}|}\right)\Big]
 +\frac{2}{3}\Bigg\rbrace \tilde{p}^\mu \tilde{p}^\nu\Delta u_{\mu\nu}. \nonumber\\
 \end{eqnarray}
The above calculations are done in the local rest frame of the 
medium. In a general frame with $4-$velocity $u^\mu$ these
results become
\begin{eqnarray}
 \frac{dN_0}{d^4x d^4p}&=&\frac{z_q^2}{2} 
 \frac{M_{eff}^2 g_d^2 \sigma(M_{eff}^2)}{(2\pi)^5} 
 e^{-u \cdot \tilde{p}/T}   \label{id_gen},\\
 \frac{dN_\chi}{d^4x d^4p}&=&\frac{dN_0}{d^4x d^4p}\tilde{p}^\mu \tilde{p}^\nu\Delta u_{\mu\nu}
 [\mathscr{P}+\mathscr{R}]
 \label{col_gen},
\end{eqnarray}
with
\begin{align}
\mathscr{P}=&\frac{2 A_q^t}{3T^3}, \nonumber \\
\mathscr{R}=&\frac{A_q^t}{T^3}\frac{\delta \omega_q}{[(u \cdot \tilde{p})^2 - M_{eff}^2]^2}
\Bigg\lbrace\frac{(u \cdot \tilde{p})}{2}[5M_{eff}^2 - 2(u \cdot \tilde{p})^2]\nonumber\\
&-\frac{3}{4}\frac{M_{eff}^4}{\sqrt{(u \cdot \tilde{p})^2 - M_{eff}^2}}
\ln\left(\frac{u\cdot \tilde{p}+\sqrt{(u \cdot \tilde{p})^2 - M_{eff}^2}}
{u\cdot \tilde{p} -\sqrt{(u \cdot \tilde{p})^2 - M_{eff}^2}}\right)\Bigg\rbrace. \nonumber
\end{align}
Next, we proceed to calculate the dilepton production rate in the presence of anomalous correction 
by switching off the collisional effects in the medium.
To study the impact of anomalous transport separately, we consider the case of $\tilde{a}_C=0$
and obtain $A^a$ following the same formalism as described in section II. In the collision-less limit ( $i.e.$, $A^t = A^a$), the 
linear perturbation of the distribution function can be defined as,
\begin{equation}
 f_{1}=\frac{{A_q^a}}{T^3}\tilde{p}^\mu \tilde{p}^\nu \Delta u_{\mu\nu},
\end{equation}
where,
\begin{equation}
 {A_q^a}=\frac{20}{9}\frac{C_fN_c T^3}{(N_c^2-1) \hat{q}}\frac{I_5^q}{I_4^q}, \label{Aq0}
\end{equation}
with $C_f=\frac{N_c^2-1}{2N_c}$ for quarks and $\hat{q}$ as the jet quenching factor. In Ref.~\cite{Majumder:2007zh}, the authors have realized that the parameter $\hat{q}$ is proportional to the mean momentum square per unit length on the particle imparted by turbulent color fields. The strength of the turbulent fields, $\langle E^2+B^2\rangle_{k}$, can be related to the jet quenching parameter as,
\begin{equation}
\hat{q}=\dfrac{2g^2C_{g/f}}{3(N_c^2-1)}\langle E^2+B^2\rangle\tau_m.
\end{equation}
Following a similar procedure, we can obtain 
the non-equilibrium contribution (without collisional effects) to the dilepton production rate as,  
\begin{eqnarray}
 \frac{dN_\chi}{d^4x d^4p}=\frac{dN_0}{d^4x d^4p}\frac{2 {A_q^a}}{3T^3}
 \tilde{p}^\mu \tilde{p}^\nu \Delta u_{\mu\nu}. \label{chi_1}
\end{eqnarray}
The total dilepton production rate for this case is obtained by adding the expressions Eq.~(\ref{id_gen}) and Eq.~(\ref{chi_1}),
\begin{eqnarray}
\frac{dN}{d^4x d^4p}=\frac{dN_0}{d^4x d^4p}\left[1+\frac{2 {A_q^a}}{3T^3}
\tilde{p}^\mu \tilde{p}^\nu \Delta u_{\mu\nu}\right]. \label{anom_gen}
\end{eqnarray}
 \begin{figure}
 \includegraphics[scale=0.5]{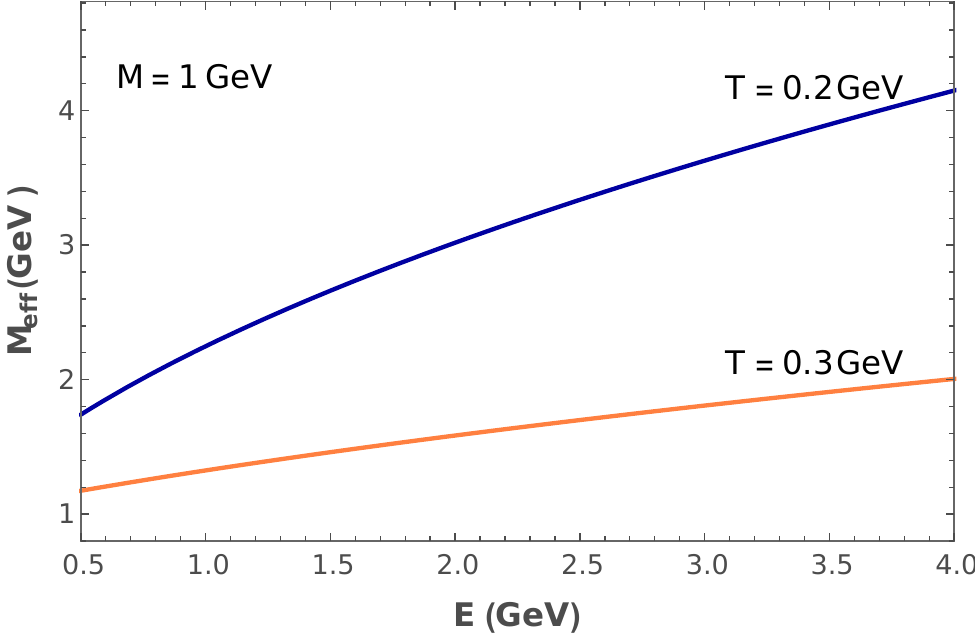}
 \caption[short]{$M_{eff}$ dependence on energy for various temperatures.}
  \label{fig:Meff_1} 
\end{figure}
Now, we write the thermal dilepton yields obtained in the limit $M_{eff}\rightarrow M$, i.e.,
when the modifications on $M$ due to the medium effects are neglected. 
We note that in this limit, the expression for $M_{eff}^2$ reduces to
$M^2 = (E_1+E_2)^2 - (\vec{p}_1 + \vec{p}_2)^2$, the invariant mass of virtual photon.
Here, $p_{1,2} = (E_{1,2},\vec{p}_{1,2})$ represents the $4-$momenta of the quark and antiquark.
Also, the $4-$momentum of dilepton reduces to $p^\mu = (p_0=E_1 + E_2 , \vec{p} =
\vec{p}_1 + \vec{p}_2)$. 
Now, within this limit, the rate of dilepton production for $q\bar{q}$ annihilation
process given by Eq. (\ref{rate_m}) takes the form~\cite{rvogt},
\begin{eqnarray} \label{dil_rate}
 \frac{dN}{d^4x d^4p} &=& \int \frac{d^3\vec{p}_1}{(2\pi)^3} \frac{d^3\vec{p}_2}{(2\pi)^3}\,
 \frac{M^2 g_d^2 \sigma(M^2)}{2 E_1 E_2}\nonumber\\
 && \times f(\vec{p}_1) f(\vec{p}_2) \delta^4(p-p_1-p_2).
\end{eqnarray}
The cross-section for this process in Born approximation is well known and with $N_f=2$, 
$N_c=3$ we have $M^2g_d^2\sigma(M^2)=\frac{80\pi}{9}\alpha^2$
~\cite{Alam:1996fd,Domokos:1980ba}. 

Within this limit, the equilibrium contribution to thermal dilepton rate has 
the form~\cite{Chandra:2015rdz}
\begin{eqnarray}
 \frac{dN_0}{d^4x d^4p}&=&\int \frac{d^3\vec{p}_1}{(2\pi)^6}\,z_q^2 e^{-(E_1+E_2)/T} 
 \frac{M^2g_d^2\sigma(M^2)}{2E_1E_2}    \nonumber\\
 &&\times\delta(p_0-E_1-E_2)    \nonumber \\
 &=&z_q^2 \times \frac{1}{2}\frac{M^2g_d^2\sigma(M^2)}{(2\pi)^5}e^{-(u\cdot p)/T}.  \label{id_gen2}
\end{eqnarray}
The non-equilibrium contributions to the dilepton production rate, with and without the collisional terms are 
obtained by taking the limit
$M_{eff}\rightarrow M$ in the Eqs. (\ref{col_gen}) and (\ref{anom_gen}) respectively.

\begin{figure}
 \includegraphics[scale=0.5]{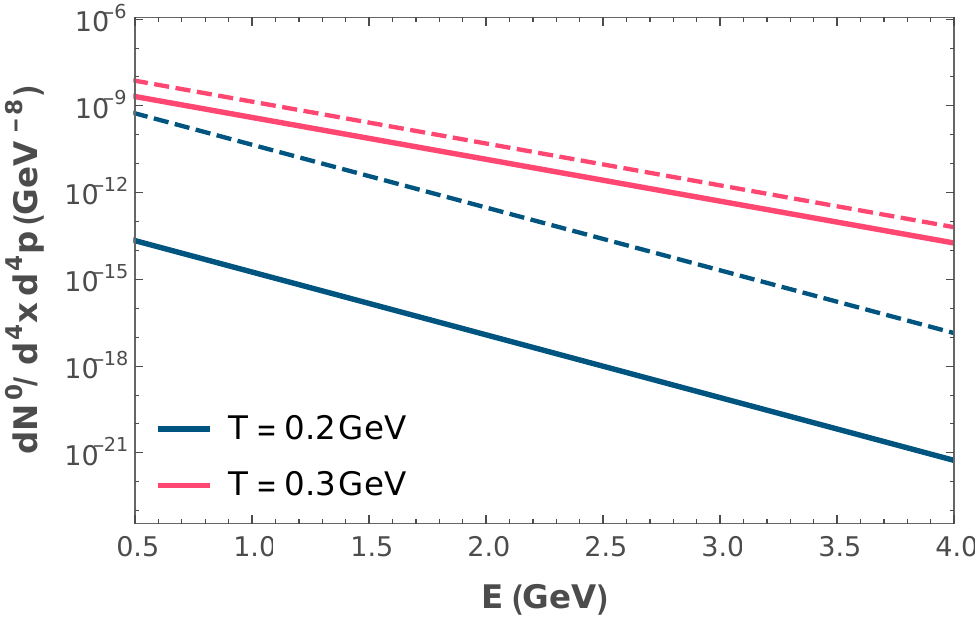}
 \caption[short]{Equilibrium dilepton production rate 
 plotted against $E$. The dashed lines represent the 
 rates in the limit $M_{eff}\rightarrow M$.} 
  \label{fig:Eq_rate1} 
\end{figure}

Now, we analyze the strength of the medium interaction effects on thermal
dilepton rates. As a first step to this analysis, we examine the temperature dependence of $M_{eff}$.
Fig.~\ref{fig:Meff_1} shows $M_{eff}$ plotted for various energies with $M=1$ GeV. 
It is evident that the impact of medium effects is more dominant at low temperatures.
Also, we note that irrespective of the value of $M$, at high temperature,
$M_{eff}$ approaches $M$, which is indicative of the fact that the interaction
term vanishes with the increase of temperature. 
In Fig.~\ref{fig:Eq_rate1}, we analyze the strength of these interaction terms 
on thermal dilepton rate by plotting the equilibrium 
dilepton production rate obtained within the quasi-particle prescription (Eq.~(\ref{id_gen2}), 
denoted by solid lines) along with the one calculated in the
limit $M_{eff}\rightarrow M$ (Eq.~(\ref{id_gen}), represented by dashed lines).
It is observed that the presence of medium interaction effects suppresses 
the dilepton rates at all energies.
This suppression is notably high at low temperatures, which 
indicates the presence of strong medium effects at low temperatures. Here, we note that the
medium interaction effects have a significant impact on the dilepton rates and hence these effects
has to be incorporated in the following analysis.
\begin{figure}
 \includegraphics[scale=0.5]{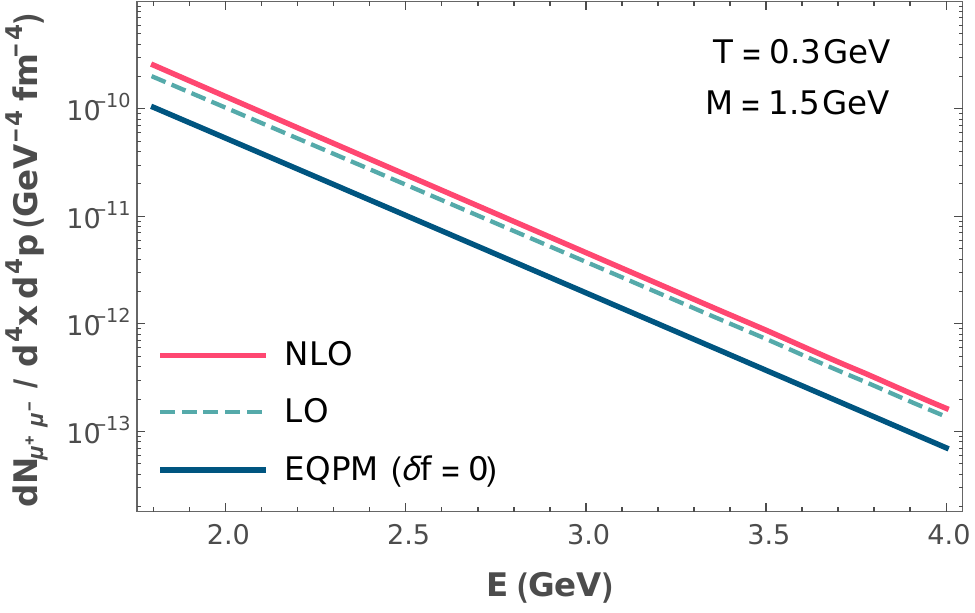}
 \caption[short]{Comparison of dilepton rate calculated within EQPM along with the strict NLO results for dilepton production. LO and NLO rates are taken from Ref.~\cite{Laine:2013vma}.}
  \label{fig:eqpm_nlo} 
\end{figure}

Note that, in the present work, we focus on the $q\bar{q}$ annihilation process of dilepton production in the Born approximation. However, there are other dilepton rate calculations in the presence of higher-order corrections~\cite{Song:2018dvf,Laine:2013vma,Laine:2015iia,Aurenche:2002wq}. In Fig.~\ref{fig:eqpm_nlo}, we show a comparison of production rate of $\mu^+ \mu^-$ pairs calculated within the EQPM (for $\delta f=0$) with the strict next-to-leading order (NLO) results of Ref~\cite{Laine:2013vma}. The leading order (LO) rate is also plotted for comparison. The rates are plotted as a function of the energy of the dilepton pair ($E$) for a fixed temperature, $T=0.3$ GeV, and invariant mass, $M=1.5$ GeV. For this analysis, we evaluate the dilepton rate given by Eq.~(\ref{dil_rate}) without considering the Maxwell-Boltzmann approximation. Also, we take $N_c=N_f =3$ for this comparison. From fig.~\ref{fig:eqpm_nlo}, it can be seen that the presence of fugacity parameter suppresses the $\mu^+ \mu^-$ rate when compared to LO results for all dilepton energies. This is in line with the result obtained in Ref.~\cite{Chandra:2016dwy}. It is also observed that while comparison with NLO results, the EQPM rate suffers more decrement compared to the previous case.

\section{Thermal dilepton yield from expanding QGP}

Dilepton yield from QGP in heavy-ion collisions can be studied by 
obtaining the temperature profile of the system. This can be done
by modelling the expansion of QGP using relativistic hydrodynamics. We 
employ the longitudinal boost invariant flow model of Bjorken to study 
the expansion of the system. In this model, the coordinates are 
parametrized as $t=\tau\cosh\eta_s$ and $z=\tau\sinh\eta_s$, 
where, $\tau=\sqrt{t^2-z^2}$, is the proper time and
$\eta_s=\frac{1}{2}\ln\left[\frac{t+z}{t-z}\right]$ is the
space-time rapidity of the system and the fluid $4-$velocity 
is expressed as $u^\mu=(\cosh\eta_s, 0, 0, \sinh\eta_s)$~\cite{Bjorken}. Now,
four dimensional volume element is given by 
$d^4x=\pi R_A^2d\eta_s\tau d\tau$, where $R_A=1.2A_a^{1/3}$ 
is the radius of the nucleus used for collision (for $Au$, $A_a=197$). 
The $4-$momentum of the dilepton can be parametrized as 
$\tilde{p}^\alpha=(M_T\cosh y, p_T \cos\phi_p, p_T\sin\phi_p, M_T\sinh y)$, 
where $M_T^2=p_T^2+M_{eff}^2$. 
Now, the factors appearing in the rate 
expression under Bjorken expansion can be calculated as
\begin{align}
 u\cdot \tilde{p}=&M_T\cosh(y-\eta_s), \label{eq1}\\
{\tilde{p}^\mu \tilde{p}^\nu \Delta u_{\mu\nu}=}&{\frac{1}{\tau}
 \left[\frac{p_T^2}{3}-\frac{2M_T^2}{3}\sinh^2(y-\eta_s)\right]} \label{eq2}.
\end{align}
We note that, when the modification on $M$ due to the medium effects are neglected, i.e., 
in the limit $M_{eff} \rightarrow M$, the expression for $\tilde{p}^\alpha$ reduces to 
$p^\alpha = (m_T\cosh y, p_T \cos\phi_p, p_T\sin\phi_p,
m_T\sinh y)$ with $m_T^2=p_T^2+M^2$. Also, within this limit, Eqs.~(\ref{eq1}) and 
(\ref{eq2}) reduces to $u \cdot p$ and $p^\mu p^\nu \Delta u_{\mu\nu}$
respectively.

Next, we write the dilepton yields 
in terms of the invariant mass $M$, transverse momentum
$p_T$ and rapidity $y$ of the dileptons produced,
\begin{eqnarray}
\frac{dN}{dM^2d^2p_Tdy}&=&\pi R_A^2 \int_{\tau_0}^{\tau_f}d\tau\,\tau 
\int_{-\infty}^{\infty}d\eta_s\,\frac{1}{2}\frac{dN}{d^4xd^4p}\nonumber\\
&& \times \left[1+\frac{2}{m_T}\cosh(y-\eta_s)\delta \omega_q\right].
\end{eqnarray}
By using Eq.~($\ref{drate_tot}$), we write the total dilepton yield as,
\begin{equation}
 \frac{dN}{dM^2 d^2p_T dy}=\frac{dN_0}{dM^2 d^2p_T dy}+\frac{dN_\chi}{dM^2 d^2p_T dy}.
\end{equation}
The equilibrium contribution to the dilepton yield is obtained as,
\begin{eqnarray} \label{dy_id} 
 \frac{dN_0}{dM^2 d^2p_T dy}&=& \mathscr{C} \int_{\tau_0}^{\tau_f} d\tau \,z_q^2 \tau 
 \int_{-\infty}^\infty d\eta_s e^{-M_T/T \cosh(y-\eta_s)} \nonumber\\
 && \times \left[1+\frac{2}{m_T}\cosh(y-\eta_s)\delta \omega_q\right],
\end{eqnarray}
where, $\mathscr{C} = \frac{\pi R_A^2}{2^2 (2\pi)^5} \frac{80 \pi}{9} \alpha^2$.

Now, the non-equilibrium contribution to the dilepton yield can be simplified as,

\begin{widetext}
\begin{align} 
\frac{dN_\chi}{dM^2 d^2p_T dy}=&\mathscr{C} \int_{\tau_0}^{\tau_f} d\tau\,z_q^2 
 \frac{A_q^t}{T^3}  \int_{-\infty}^{\infty} d\eta_s e^{-M_T/T \cosh(y-\eta_s)}  \left[1+\frac{2}{m_T}\cosh(y-\eta_s)\delta \omega_q\right] \nonumber\\
 & \times {\left[\frac{p_T^2}{3}-\frac{2M_T^2}{3}\sinh^2(y-\eta_s)\right] \left\{\frac{2}{3} - \mathscr{E}(T,\eta_s) \,\right\}},\label{dy_col} 
\end{align}
with
\begin{align}
\mathscr{E}(T,\eta_s)=&  \frac{\delta \omega_q}{[M_T^2 \cosh^2(y-\eta_s) - M_{eff}^2]^2}
 \Bigg\{\left[M_T^2 \cosh^2(y-\eta_s) - \frac{5}{2}M_{eff}^2  \right] M_T\cosh(y-\eta_s)\nonumber\\
 &+\frac{3}{4} \frac{M_{eff}^4}{\sqrt{M_T^2 \cosh^2(y-\eta_s) - M_{eff}^2}}
 \ln{\left(\frac{ M_T \cosh(y-\eta_s) +\sqrt{M_T^2 \cosh^2(y-\eta_s) - M_{eff}^2}}
 { M_T \cosh(y-\eta_s) - \sqrt{M_T^2 \cosh^2(y-\eta_s) - M_{eff}^2}}\right)}
  \Bigg\}. \label{mathcal_E1}
\end{align}
\end{widetext}
The total dilepton yield in the presence of collisional terms can be calculated
by numerically integrating the expressions Eq.~($\ref{dy_id}$) and Eq.~($\ref{dy_col}$)
along with the temperature profile of the expanding plasma.

Further, for comparison, we calculate the dilepton yield without the collisional 
correction term. From Eq.~($\ref{anom_gen}$), the non-equilibrium contribution to the 
dilepton yield for this case is obtained as,
\begin{eqnarray}
 \frac{dN_\chi}{dM^2d^2p_Tdy}&=&\frac{2\mathscr{C}}{3}\int_{\tau_0}^{\tau_f}d\tau\,z_q^2
 \frac{{A_q^a}}{T^3} \int_{-\infty}^{\infty} d\eta_s\,e^{-M_T/T \cosh(y-\eta_s)}\nonumber \\
 &&\times \Bigg\{
 \left[1+\frac{2}{m_T}\cosh(y-\eta_s)\delta \omega_q\right] \nonumber\\
 &&\times{\left[\frac{p_T^2}{3}-\frac{2M_T^2}{3}\sinh^2(y-\eta_s)\right]} \Bigg \}, 
\end{eqnarray}
where ${A_q^a}$ is defined in Eq.~($\ref{Aq0}$).

Next, we write the thermal dilepton yields calculated within the 
limit $M_{eff} \rightarrow M$,
\begin{equation}
 \frac{dN}{dM^2d^2p_Tdy}=\pi R_A^2 \int_{\tau_0}^{\tau_f}d\tau\,\tau 
\int_{-\infty}^{\infty}d\eta_s\,\frac{1}{2}\frac{dN}{d^4xd^4p}.
\end{equation}

The equilibrium 
contribution to dilepton yield for this case can be written as
\begin{eqnarray} 
 \frac{dN_0}{dM^2 d^2p_T dy} &=& \mathscr{C} \int_{\tau_0}^{\tau_f} d\tau \,z_q^2 \tau 
 \int_{-\infty}^{\infty} d\eta_s e^{-m_T/T \cosh(y-\eta_s)} \nonumber\\
 &=& 2 \mathscr{C} \int_{\tau_0}^{\tau_f} d\tau \,z_q^2 \tau  K_0(m_T/T),
\end{eqnarray}
where $K_n$ is the modified Bessel function of second kind.
Now, the non-equilibrium contribution to dilepton yield in this limit can be obtained as
\begin{eqnarray}
 \frac{dN_\chi}{dM^2d^2p_Tdy} &=& \mathscr{C} \int_{\tau_0}^{\tau_f} d\tau z_q^2
 \frac{A_q^t}{T^3} {\Bigg\{\mathscr{T}(T)}  \nonumber\\
 &&{  -\int_{-\infty}^{\infty}d\eta_s   
  \delta \omega_q \,
 \mathscr{E}(T,\eta_s)\,\Bigg\}}, 
\end{eqnarray}
with
\begin{equation}\label{mathscr_T}
 {\mathscr{T}(T)=\frac{4}{9}\left[p_T^2 K_0(m_T/T) - 2T m_T K_1(m_T/T)  \right],}
\end{equation}
\begin{widetext}
\begin{eqnarray}
\mathscr{E}(T,\eta_s)&=&  \frac{ e^{-m_T/T \cosh(y-\eta_s)}}{[m_T^2 \cosh^2(y-\eta_s) - M^2]^2} {\left[\frac{p_T^2}{3}-\frac{2M_T^2}{3}\sinh^2(y-\eta_s)\right]}
 \Bigg\{\left[ m_T^2 \cosh^2(y-\eta_s)  - \frac{5}{2}M^2 \right] m_T\cosh(y-\eta_s)\nonumber\\
 &&+\frac{3}{4} \frac{M^4}{\sqrt{m_T^2 \cosh^2(y-\eta_s) - M^2}}
 \ln{\left(\frac{ m_T \cosh(y-\eta_s) +\sqrt{m_T^2 \cosh^2(y-\eta_s) - M^2}}
 { m_T \cosh(y-\eta_s) - \sqrt{m_T^2 \cosh^2(y-\eta_s) - M^2}}\right)}
  \Bigg\}. \label{mathcal_E2}
\end{eqnarray}
\end{widetext}

The non-equilibrium contribution to the yield without collisional effects calculated within the limit $M_{eff} \rightarrow M$ is given by
\begin{eqnarray}
 \frac{dN_\chi}{dM^2d^2p_Tdy} &=& \mathscr{C} \int_{\tau_0}^{\tau_f} d\tau z_q^2 
 \frac{{A_q^a}}{T^3} \mathscr{T}(T),
\end{eqnarray}
where ${A_q^a}$ and $\mathscr{T}(T)$ are defined in Eqs.~(\ref{Aq0}) and (\ref{mathscr_T})
respectively.

\section{Results and discussions}

We obtain the temperature profile of the system by solving the 
hydrodynamical equations with initial conditions relevant to RHIC energies.
The initial time and temperature are taken to be $\tau_0=0.5\,fm/c$ 
and $T_0=300\,MeV$ respectively. The energy density evolution equation governing 
the longitudinal expansion of the plasma is given by~\cite{Bjorken}
$\frac{d\epsilon}{d\tau}+\frac{\epsilon+P}{\tau}=0$. We chose the recent 
lattice QCD EoS~\cite{cheng} in the current analysis.
The above hydrodynamic equation is solved to obtain $T(\tau)$; and we note that the system 
reaches the critical temperature $T_c$ at a time $\tau_f=5.4\,fm/c$. 
Now, we calculate the dilepton yields 
by numerically integrating the rate expressions obtained in the previous section 
with the temperature profile $T(\tau)$. We carry out the integration from 
$\tau_0$ to $\tau_f$. The yields are presented for the midrapidity region of 
the dileptons, $i.e.$, for $y=0$.  

\begin{figure}
 \includegraphics[scale=0.5]{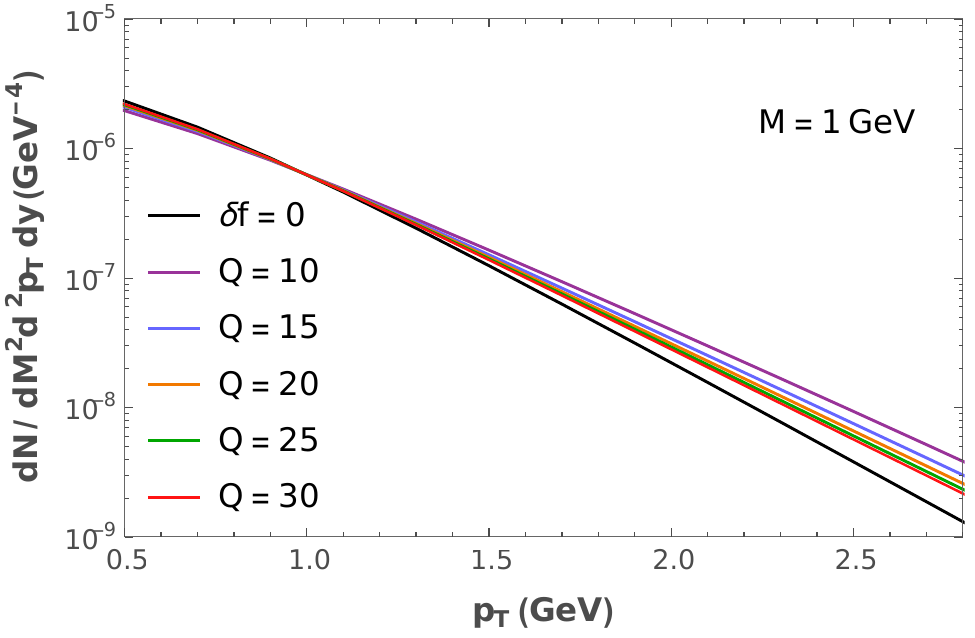}
 \caption[short]{Thermal dilepton yields in the presence of 
  collisional terms for different values of $Q\equiv\hat{q}/{T^3}$ and
  for invariant mass $M=1$ GeV. Equilibrium contribution to the dilepton 
  yield ($\delta f=0$) is also plotted for comparison.} 
  \label{fig:M1all} 
\end{figure}
Fig.~\ref{fig:M1all} shows the dilepton yields in presence of collisional terms as a 
function of transverse momentum $p_T$ for $M=1$ GeV. The yields are plotted for 
different values of the jet-quenching parameter,
$\hat{q}/T^3\equiv Q$. {It is observed that the presence of collisional
terms increases the dilepton yield considerably, compared to the equilibrium case
(represented by $\delta f = 0$). 
Note that, with $Q=10$, there is 
an increase of $\sim 31.8\%$ at $p_T=1.5$ GeV and $\sim145\%$ at 
$p_T=2.5$ GeV for $M=1$ GeV. 
It can be noted that the 
increment due to collisional terms 
decreases with the increase of $Q$. For $M=1$ GeV and
at $p_T=2$ GeV, we observe $\sim46\%$ enhancement in the yield with
$Q=20$ and $\sim30\%$ with $Q=30$. We observe that the effect of collisional terms is more 
prominent in the high $p_T$ regime, which indicates that these 
non-equilibrium effects are more dominant at high $p_T$. This is 
due to the fact that high $p_T$ particles are produced predominantly 
during the initial stages of QGP evolution. In the case of lower $p_T$, we observe a marginal 
decrease in the yields, which is indicative of 
the fact that these effects remain significant throughout the 
evolution of the plasma.}

\begin{figure}
 \includegraphics[scale=0.5]{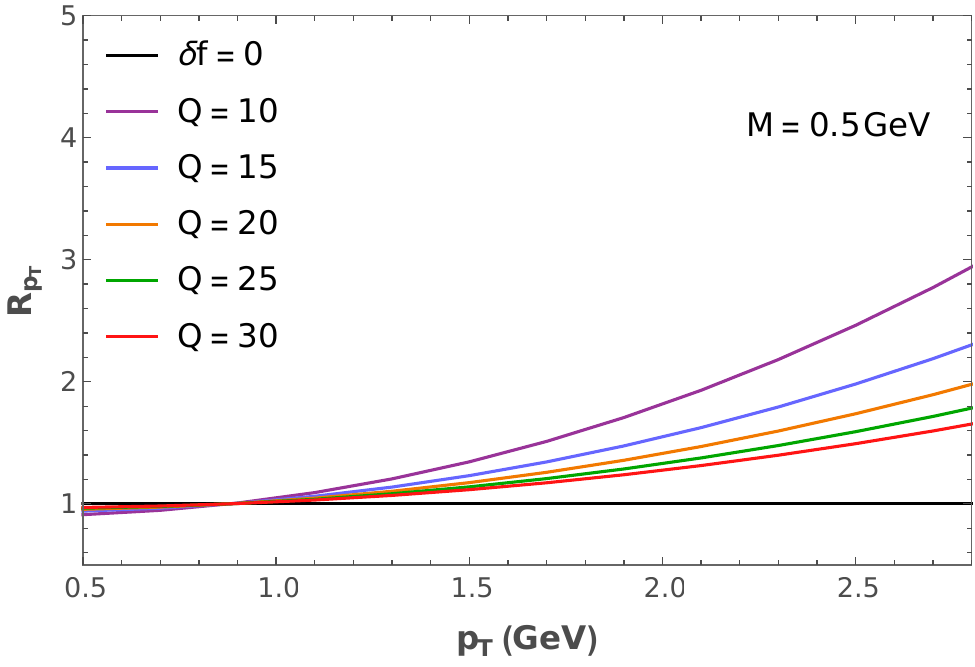}
  \caption[short]{Strength of collisional terms to the equilibrium
  dilepton yield for various values of $Q$ with invariant
  mass $M=0.5$ GeV.}
\label{fig:RptM05}
  \end{figure}

Now, we study the strength of these collisional effects to the 
equilibrium dilepton yield by constructing a ratio as given below,
 \begin{equation}
  R_{p_T}=\left[1+\frac{dN_\chi}{dM^2d^2p_Tdy}/\frac{dN_0}{dM^2d^2p_Tdy}\right].
 \end{equation}
Fig.~\ref{fig:RptM05} shows $R_{p_T}$ as a function of transverse
momentum for various values of $Q$. { We observe that the strength of collisional contributions are higher at large $p_T$ compared to
lower $p_T$. As expected, we see a gradual increase in the collisional effects 
as we move towards high $p_T$. This trend remains the same for all values 
of $M$. It is observed that the  strength of collisional corrections decreases with increase in $Q$. At low $p_T$, $R_{p_T}$ is less than unity, which indicates that the corrections suppress the particle spectra at lower $p_T$ and the maximum suppression can be seen for $Q=10$ and minimum for $Q=30$. 
}

\begin{figure}
 \includegraphics[scale=0.5]{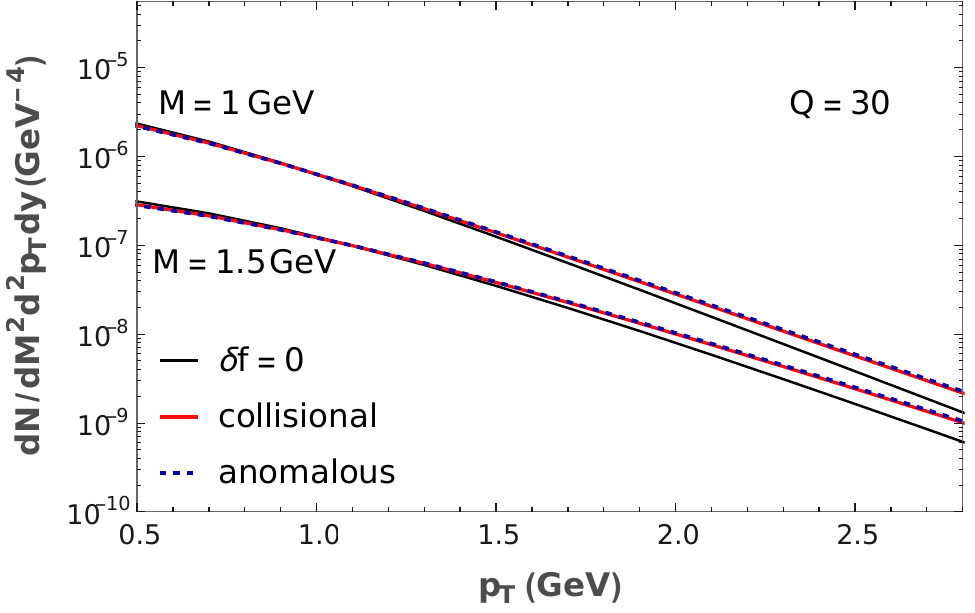}
 \caption[short]{Thermal dilepton yields in the presence of 
  collisional and anomalous corrections with $Q=30$, for different values of $M$.} 
  \label{fig:Mplot} 
\end{figure}
  
Next, we compare the dilepton yields obtained for collisional and anomalous corrections 
in Fig.~\ref{fig:Mplot}. 
In doing so, we plot the yields for $M=1, 1.5$ GeV while fixing $Q=30$. 
{Though the strength of non-equilibrium collisional effects on spectra is appreciable at high $p_T$,
it is found to be lesser compared to that of the anomalous transport. 
We note that, over entire $p_T$, the effect due to anomalous transport is more compared to the collisional case.
At $p_T=2.5$ GeV and with $M=1$ GeV, we observe $\sim48.8\%$ increase in the yield in the presence of collisional effects, while for the same parameters the increment is $\sim54.8\%$ for the anomalous case.
}

\begin{figure}
 \includegraphics[scale=0.5]{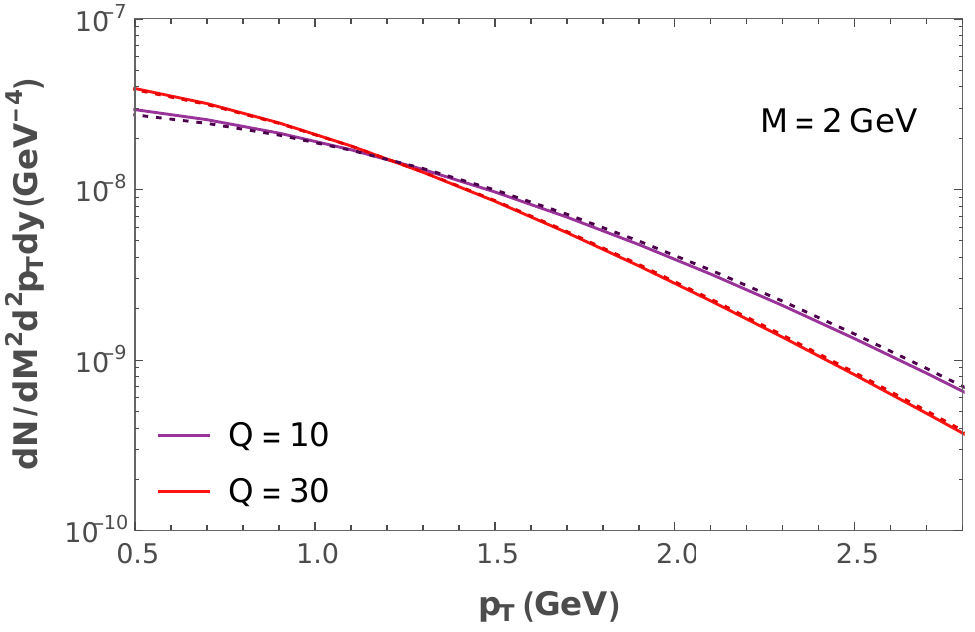}
 \caption[short]{Thermal dilepton yields in the presence of 
  collisional terms for different values of $Q\equiv\hat{q}/{T^3}$ and
  for invariant mass $M=2$ GeV. The dotted lines indicate yields 
  from the anomalous transport only.} 
  \label{fig:M2plot} 
\end{figure}

{Fig.~\ref{fig:M2plot} shows the 
dilepton yields for collisional and anomalous corrections as a 
function of transverse momentum for $M=2$ GeV with different $Q$ values. 
Though, the effect of both the corrections is to increase the equilibrium dilepton yield at large $p_T$, the enhancement is found to be lesser when collisional terms are included. 
It is to be noted that difference between the two corrections
is visibly observed for high $p_T$ and small $Q$. 
Our analysis indicates that the dilepton yield in the presence of collisional terms is lesser when 
compared to the collisionless anomalous transport case.} 
{This is in line with the 
argument of Ref. \cite{Asakawa:2006jn}
that the $2\rightarrow2$ elastic collisions have only marginal contributions 
to transport coefficients as 
compared to that from the turbulent chromo fields described through effective 
Vlasov-Boltzmann equation.}

\begin{figure}
 \includegraphics[scale=0.5]{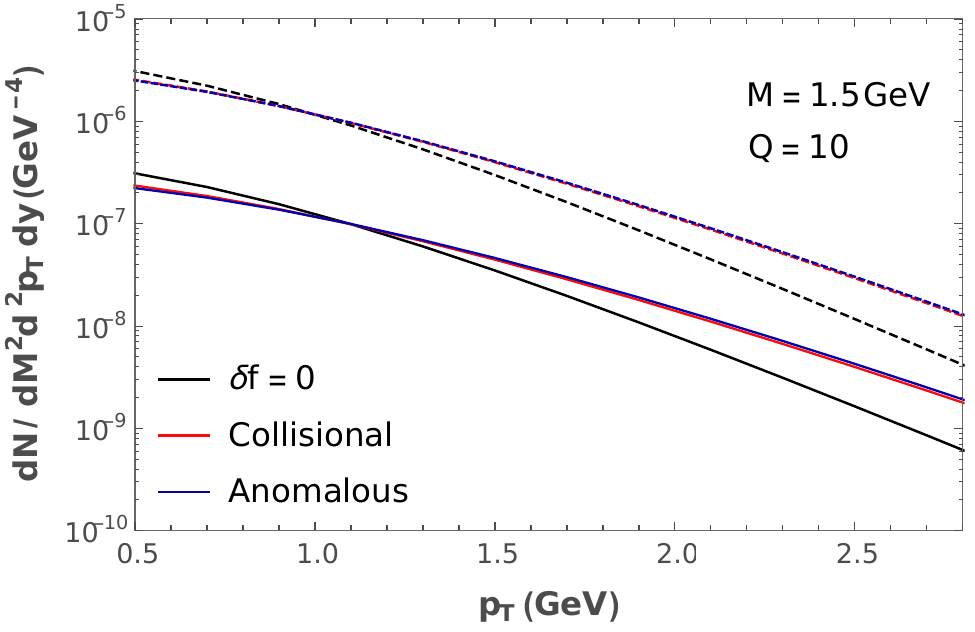}
 \caption[short]{The effect of medium modified mass 
 $M_{eff}$ on the thermal dilepton yields in the presence of 
  collisional and anomalous corrections. The yields are plotted for $Q=10$ and
  $M=1.5$ GeV. The dashed lines indicate the 
  yields obtained for the case $M_{eff} \rightarrow M$.} 
  \label{fig:rate1} 
\end{figure}

{In Fig.~\ref{fig:rate1}, we plot the thermal dilepton yields in the presence of
collisional and anomalous corrections observed for $M=1.5$ GeV and $Q=10$.
Corresponding yields obtained in the $M_{eff} \rightarrow M$ limit are also 
plotted for comparison. As expected, we observe that the medium
interaction effects suppress the spectra throughout the entire $p_T$ range compared
to the $M_{eff} \rightarrow M$ limit. 
It can be seen that the difference between collisional and anomalous corrections at large $p_T$ is more visible when medium interaction effects are included. As the strength of momentum anisotropy varies with the evolution of medium, our results have a strong dependence on the temperature of the medium, time scale of instability in the medium and the choice of jet quenching parameter.}
It must be noted that the results presented here incorporate the effects of medium interactions
on the cross-section, whereas in Ref.~\cite{Chandra:2016dwy},
these effects were not considered while calculating the dilepton spectra in the presence 
of anomalous transport. 

In the present 
analysis, the collisional corrections to the thermal dilepton 
spectra are calculated using $(1+1)-$D Bjorken flow. It is to be noted that, in general, the 
Bjorken model tends to overestimate the particle production yields as the evolution time 
of the QGP is high compared to a three-dimensional flow. 
Also, we have not incorporated Debye screening corrections to thermal dilepton rates, since 
its effect is found to be minimal~\cite{Chatterjee:1994dq} in the current analysis. 
A quantitative study of collisional term correction to the spectra can be done by employing 
a $(2+1)-$D hydrodynamic flow and also including contributions from radiative processes/inelastic collisions.   {
Moreover, apart from the dominant source considered, there are other higher-order processes that can also 
contribute to the thermal dilepton production~\cite{Altherr:1992th,Thoma:1997dk,Burnier:2015rka,Jackson:2019yao}}. It would be interesting
to incorporate contributions from such channels along with the collisional 
corrections.
This will be taken up for explorations in the near future.

\section{Conclusion and outlook}

In conclusion, we have estimated the thermal dilepton production rate while incorporating 
the collisional effects of the QGP medium along with the anomalous contributions. 
We have 
employed an effective Vlasov-Boltzmann equation to 
describe the dynamics of the medium in the presence of turbulent fields.  
The Vlasov term of the transport equation describes the evolution of distribution function with turbulent chromo-fields, whereas the collision kernel quantifies the effects of collisional processes in the rate of 
change of distribution function.  
We have analyzed the effect of these non-equilibrium corrections in thermal dilepton production from $q\overline{q}$ annihilation. The effects of the collisional processes in the presence of turbulent fields to the 
dilepton production rate are quantified in the case of $(1+1)-$D boost invariant 
expansion of the medium in the heavy-ion collision scenario.

The non-equilibrium effects are found to have a visible impact on the dilepton 
spectra. {The effect of collisional corrections is to enhance the yield at high $p_T$, while it suppress the equilibrium 
dilepton spectra at lower $p_T$.}
Collisional effects in the dilepton production rate and yield are seen to 
have a strong dependence on the jet-quenching parameter ($Q \equiv \hat{q}/T^3$). 
{Notably, the enhancement to the spectra is large for small $Q$ values. }
Further, we have analyzed the dependence of invariant mass $M$ 
to the collisional corrections to the dilepton production rate. 
In addition to this, the interplay of collisional processes and anomalous transport in the 
 QGP medium is analysed through its strength on the dilepton production rates.
{The inclusion of collisional terms in the presence of chromo-turbulent 
fields suppressed the yield contribution from collisionless anomalous transport; 
and the difference is found to be more prominent in the high $p_T$ regime of the spectra.}
 Moreover, we have analyzed the effects of medium interactions on the cross-section 
 and studied its impact on 
 thermal dilepton spectra in the presence of both collisional and anomalous corrections. The inclusion of medium effects has a significant impact on the yields, and it is 
 found to suppress the dilepton spectra throughout the
 entire $p_T$ regime.

We intend to study the impact of collisional processes with both shear and bulk 
viscous effects on thermal dilepton spectra in heavy-ion collisions by 
employing a $(2+1)$-D hydrodynamical expansion of the system in the near future. 
Investigating the dilepton production rate in the magnetized QGP is another interesting 
direction to focus while utilizing the effective models for hot magnetized QCD medium~\cite{Kurian:2017yxj}. 
We leave these aspects for future works.

\section*{ACKNOWLEDGMENTS}
 V.S. would like to thank the warm hospitality of IIT Gandhinagar during his visit, 
where this work was initiated. {L.J.N. acknowledges the Department of Science and Technology, Government of India for the INSPIRE Fellowship.}
We record our gratitude to the people of India for their generous support 
for the research in basic sciences. {The authors would like to thank the anonymous referees
of this article for comments which led to significant improvement of the manuscript.}


\end{document}